\documentclass[prb,twocolumn,amsmath,amssymb]{revtex4}

\usepackage{graphicx}
\usepackage{dcolumn}
\usepackage{bm}

\begin{document}

\title{Enhancement of rare-earth--transition-metal exchange interaction in Pr$_{2}$Fe$_{17}$
probed by inelastic neutron scattering}

\author{N. Magnani}
\author{S. Carretta}
\author{G. Amoretti}
\affiliation{Istituto Nazionale per la Fisica della Materia, Universit\`a di Parma,
Parco Area delle Scienze 7/A, I-43100 Parma, Italy}

\author{L. Pareti}
\author{A. Paoluzi}
\affiliation{Istituto I.M.E.M. del Consiglio Nazionale delle Ricerche,
Parco Area delle Scienze 37/A, I-43010 Fontanini (PR), Italy}

\author{R. Caciuffo}
\affiliation{Istituto Nazionale per la Fisica della Materia, Universit\`a
Politecnica delle Marche, Via Brecce Bianche, I-60131 Ancona, Italy}

\author{J. A. Stride}
\affiliation{Institut Laue-Langevin, Boite Postale 220 X, F-38042 Grenoble
Cedex, France}

\date{\today}

\begin{abstract}
The fundamental magnetic interactions of Pr$_{2}$Fe$_{17}$ are
studied by inelastic neutron scattering and anisotropy field
measurements. Data analysis confirms the presence of three
magnetically inequivalent sites, and reveals an exceptionally
large value of the exchange field. The unexpected importance of
$J$-mixing effects in the description of the ground-state
properties of Pr$_{2}$Fe$_{17}$ is evidenced, and possible
applications of related compounds are envisaged.
\end{abstract}

\maketitle

Rare-earth--transition-metal (RE--TM) intermetallic compounds have
been extensively studied during the last decades,\cite{bus91}
allowing to design and produce high-performance permanent magnets
for industrial use such as Sm$_{2}$Co$_{17}$ and
Nd$_{2}$Fe$_{14}$B. The mean-field Hamiltonian which describes the
RE quantum state in these compounds can be written as
\begin{equation}
\hat{H}_{RE}=\Lambda {\bf \hat{L}\cdot \hat{S}}+2\mu_{B}{\bf
H_{ex}\cdot \hat{S}} +\sum_{k,q} B_{kq} \hat{C}_{q}^{(k)},
\label{ham}
\end{equation}
where the first term on the right-hand side is the spin-orbit
coupling, the second is the exchange interaction treated in
mean-field and the third is the crystal field (CF). It has been
shown that the leading anisotropy constant $K_{1}$ is
approximately proportional to $\alpha B_{20}H_{ex}^{2}$ at high
temperatures (where $\alpha$ is the second-order Stevens factor
for the considered RE ion);\cite{kuz95} therefore, as a general
rule, the exchange field $H_{ex}$ should be large and the product
$\alpha B_{20}$ must be negative in order to obtain the high
easy-axis anisotropy required to make good permanent magnets.
While much work was made to tailor the CF potential as needed
(which led to the discovery that, in some cases, the insertion of
small amounts of interstitial nitrogen or carbon gives rise to a
strong easy-axis anisotropy\cite{bus91}), fewer efforts have been
devoted to study the exchange interaction, which is essential in
determining the magnetic behavior of the RE sublattice. In
particular, while a large enhancement of the RE-TM exchange has
been envisaged going from heavier towards lighter rare-earths (due
to the different spatial extent of the $4f$ electronic
wavefunctions), only indirect experimental evidences are available
({\it i.e.} the exchange constant $n_{RT}$ is determined by
measurements of the Curie temperature $T_{C}$).\cite{bel87}

\begin{figure}
\includegraphics[width=85mm]{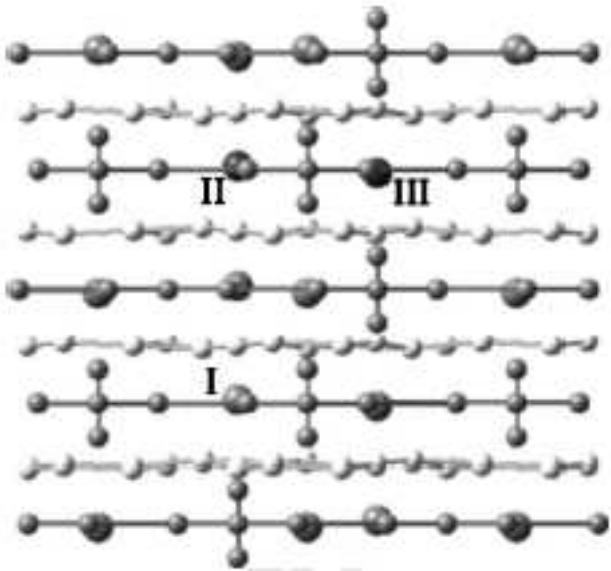}
\caption{\label{fig:structure} Crystallographic structure of
Pr$_{2}$Fe$_{17}$ (after Ref.~\onlinecite{cal03}). Large dark
spheres are Pr atoms, while the smaller spheres representing Fe
atoms are indicated in two different colors (darker for those
atoms belonging to a Pr-Fe plane, lighter for those belonging to a
Fe-only plane). The three inequivalent RE sites are indicated by
Roman numerals.}
\end{figure}

Since neutrons are excellent probes of the microscopic
magnetization, inelastic neutron scattering (INS) offers the most
direct and reliable experimental method for the determination of
CF and exchange interactions in rare-earth
intermetallics.\cite{moze,til99,isnard,kuz02} The eigenstates of
Eq.~(\ref{ham}) are mainly determined by the exchange term, so
they can be labelled by quantum numbers as $\left| J,M \right>$;
at low temperature, the selection rules for magnetic dipole allow
only one intramultiplet transition ($\left| J,-J \right>
\rightarrow \left| J,-J+1 \right>$) from the ground state. In the
present work, INS experiments have been performed on
Pr$_{2}$Fe$_{17}$. Recent investigations have definitely put into
question the apparently simple structural features of this
compound;\cite{cal03} in particular, the unusual frequency of
obverse-reverse twinning results in the presence of three distinct
RE sites, which are not equivalent from the crystallographic point
of view (Fig.~\ref{fig:structure}). Multiple competing
contributions to the anisotropy have been identified as one of the
sources of the complex magnetic phase diagram and transitions
observed for the Pr$_{2}$(Fe$_{1-x}$Co$_{x}$)$_{17}$
series.\cite{par02,kou98} Moreover, a INS determination of the
exchange interaction for Sm$_{2}$Fe$_{17}$ has been recently
published,\cite{sol02} allowing for a direct comparison of the
results.

\begin{figure}
\includegraphics[width=85mm]{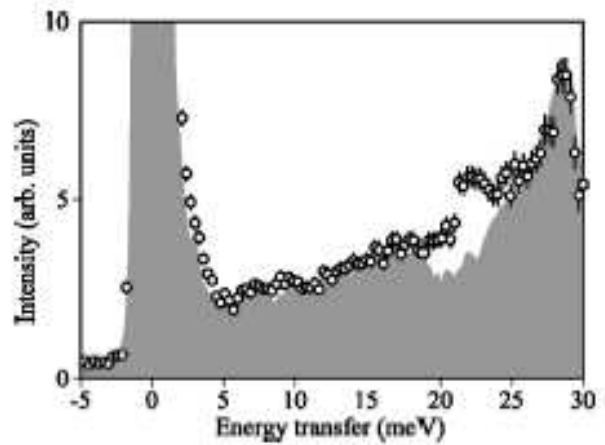}
\caption{\label{fig:PrYspec} Dots: low-angle INS spectra of
Pr$_{2}$Fe$_{17}$. Grey area: phonon background, as estimated by
Y$_{2}$Fe$_{17}$ measurements.}
\end{figure}

Polycrystalline samples of Pr$_{2}$Fe$_{17}$ and Y$_{2}$Fe$_{17}$
were obtained from high-purity (99.99\%) elements by arc-melting
technique in a water-cooled copper crucible under Ar pressure. The
ingots were remelted three times to insure homogeneity, wrapped in
Ta foil, annealed under Ar atmosphere at 950$^{{\rm o}}$C for
three days, and quenched in water; the samples were then
hand-crushed into fine powders, under protective atmosphere.
Thermomagnetic analysis\cite{par02} and X-ray diffraction showed
the presence of the 2:17 phase only. Inelastic neutron scattering
(INS) experiments for both compounds were performed on the IN4
time-of-flight spectrometer at the Institut Laue-Langevin; in
order to detect the intramultiplet excitation, the incident energy
value was fixed at 38 meV. The low-scattering-angle spectra
obtained for both compounds are shown in Fig.~\ref{fig:PrYspec}.
The data collected on Y$_{2}$Fe$_{17}$ were used to estimate the
phonon scattering. This was properly renormalized and subtracted
to the INS spectra of Pr$_{2}$Fe$_{17}$ in order to obtain the
pure magnetic contribution, which is found to be significant only
between 19 and 24 meV; its almost non-dispersive
character\cite{cla82} was checked by examining several scans taken
at different $Q$ values.

\begin{figure}
\includegraphics[width=85mm]{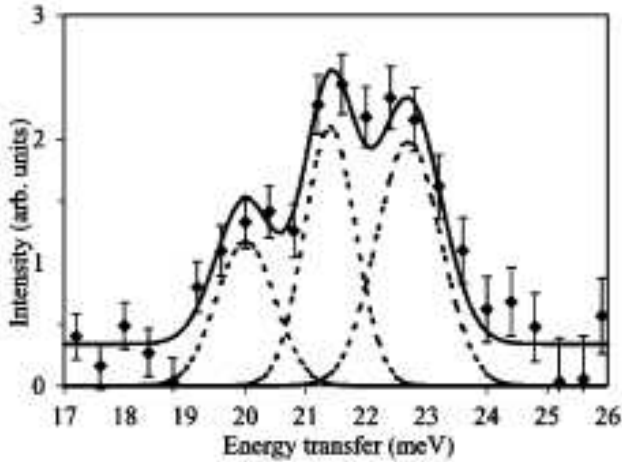}
\caption{\label{fig:magspec} INS spectrum of Pr$_{2}$Fe$_{17}$.
All the nonmagnetic contributions have been estimated and
subtracted by analyzing the corresponding Y$_{2}$Fe$_{17}$
measurements. The solid line is a fit with three gaussians,
centered respectively at $20.0$, $21.4$, and $22.8$ meV.}
\end{figure}

The fitting results, shown in Fig.~\ref{fig:magspec}, display the
presence of three equally spaced peaks. In fact, it is
straightforward to prove that the three crystallographically
inequivalent sites are also expected to be different from the
point of view of fundamental magnetic interactions by working out
an estimate of the relative exchange-field strength. Considering
the Hamiltonian
\begin{equation}
\hat{H}_{exchange}=-2\sum_{T}J_{RT}{\bf \hat{S}}_{R}\cdot {\bf
\hat{S}}_{T} \label{exchange}
\end{equation}
which describes the RE-TM exchange interaction and using a
mean-field model,\cite{bus91} one finds that the molecular field
experienced by the RE moment (which, in turn, is proportional to
the so-called exchange field ${\bf H}_{ex}$) is $n_{RT}{\bf
M}_{T}$, where ${\bf M}_{T}$ is the TM sublattice magnetization
and
\begin{equation}
n_{RT} = \sum_{T} J_{RT} \simeq zJ_{RT}
\end{equation}
where $z$, the number of nearest Fe neighbors, is different for
each site (19 for site I, 18 for site II, and 20 for site III).
Hence,
\begin{equation}
H_{ex}^{(I)}=\frac{19}{18}H_{ex}^{(II)}=\frac{19}{20}H_{ex}^{(III)}.
\label{nrt}
\end{equation}
Each detected peak should then correspond to the allowed
intramultiplet transition of one RE site. The obtained linewidths
are equal (for the central peak) or very slightly larger (for the
two side peaks) than the instrumental resolution; this line
broadening may be due to crystallographic disorder (not uncommon
for 2:17 phases\cite{moz94}) or to a weak dispersion of the
localized modes.\cite{col89} The relative transition intensities
are roughly consistent with the estimated twinning volume fraction
determined by X-ray diffraction.\cite{cal03}

The energy $\Delta$ of the intramultiplet transition corresponding
to a single RE site can be written as\cite{kuz02}
\[
\Delta = \Delta_{ex} + \Delta_{CF} \simeq 2\mu_{B}H_{ex}\times
\bigg[ \frac{1}{5} - \frac{84\mu_{B}H_{ex}}{1375\Lambda}
\]
\begin{equation}
+\frac{84}{34375}\left(
\frac{2\mu_{B}H_{ex}}{\Lambda}\right)^{2}\bigg]
-\frac{91}{825}B_{20}+\Delta_{4,6}
    \label{delta}
\end{equation}
where the dominant terms proportional to $H_{ex}$ and $B_{20}$
have been explicitly separated from the smaller contribution of
the fourth- and sixth-rank CF parameters. The latter was estimated
as $\Delta_{4,6}=3.7 \pm 2.5~{\rm meV}$, using an average of the
literature parameters for other $R_{2}$Fe$_{17}$ compounds
($R$~=~Sm, Dy, Ho, Er)\cite{rfe1,rfe2,rfe3} after rescaling them
to account for the different $4f$-wavefunction radii between
Pr$^{3+}$ and other ions ($B_{40}=-330 \pm 80$ K; $B_{60}=40\pm
60$ K; $B_{66}=-410 \pm 100$ K). It must be noticed that $J$
mixing terms are included up to the second order in $1/\Lambda$.

The correct values of $H_{ex}$ and $B_{20}$ cannot be
simultaneously obtained with Eq.~(\ref{delta}) alone; however,
another independent equation linking these parameters can be
derived if the temperature dependence of the anisotropy field
$H_{A}$ is known. The second-order anisotropy constant can be
obtained by the formula $K_{1}=H_{A}M_{S}/2$ and, just below
$T_{C}$,\cite{kuz02b,mag03}
\[
K_{1}=-\frac{11}{25}\alpha B_{20}\left(2\mu_{B}H_{ex} \right)^{2}
\]
\begin{equation}
\times \left[ \frac{7}{2}(k_{B}T)^{-2} +
\frac{6}{\Delta_{SO}}(k_{B}T)^{-1} \right]. \label{k1b20}
\end{equation}
where the spin-orbit gap is $\Delta_{SO}=267~{\rm
meV}$.\cite{kuz02b} Again, $J$-mixing is taken into account
perturbatively.\cite{mag03} Although bulk techniques such as
magnetization and anisotropy field measurements do not allow to
separate the contribution of different sites, contrarily to
spectroscopic data, it can be proved that Eq.~(\ref{k1b20})
remains valid if one replaces $B_{20}$ and $H_{ex}$ with their
weighted averages
\begin{equation}
\tilde{B_{20}}=c_{(I)}B_{20}^{(I)}+c_{(II)}B_{20}^{(II)}+c_{(III)}B_{20}^{(III)}
\end{equation}
and
\begin{equation}
\tilde{H_{ex}}=c_{(I)}H_{ex}^{(I)}+c_{(II)}H_{ex}^{(II)}+c_{(III)}H_{ex}^{(III)},
\end{equation}
where $c_{(i)}$ is the relative abundance of site $i$ within the
crystal; moreover, from Eq.~(\ref{nrt}) and considering that
$c_{(II)}=c_{(III)}$,\cite{cal03} one immediately has
$\tilde{H_{ex}}=H_{ex}^{(I)}$.\cite{nota1}

\begin{figure}
\includegraphics[width=85mm]{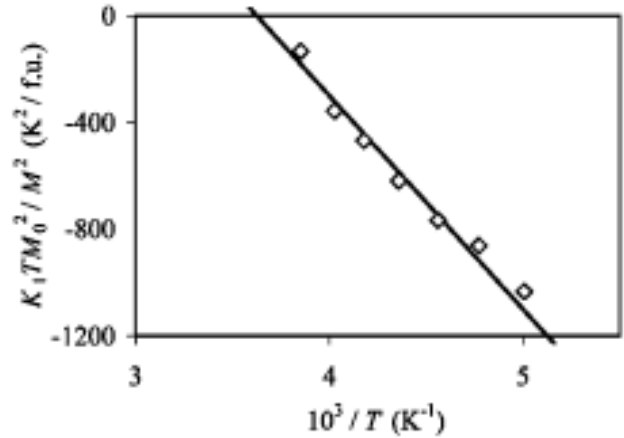}
\caption{\label{fig:results} Linear dependence of
$K_{1}T[M_{S}(0)/M_{S}(T)]^{2}$ as a function of $1/T$ above 200
K.}
\end{figure}

Anisotropy-field measurements have been performed on oriented
Pr$_{2}$Fe$_{17}$ and Y$_{2}$Fe$_{17}$ powders by the Singular
Point Detection (SPD) technique;\cite{ast74} the results for the
Pr compound are consistent with those published by Kou {\it et
al.} (Ref.~\onlinecite{kou98}). The data for the Y compound have
been used as a `blank' in order to estimate and subtract the
$3d$-electron contribution to $K_{1}$. Only the anisotropy field
data above 200 K were considered in order to be sure of the
validity of Eq.~(\ref{k1b20}). Assuming that the temperature
dependence of $H_{ex}$ follows that of $M_{S}$,\cite{sol02} a
linear dependence of $K_{1}T[M_{S}(0)/M_{S}(T)]^{2}$ as a function
of $1/T$ is actually found (Fig.~\ref{fig:results}), and the
linear fit gave the value
$\tilde{B_{20}}\left(2\mu_{B}H_{ex}^{(I)}
\right)^{2}k_{B}^{-3}\simeq -2.47 \times 10^{7}~{\rm K}^{3}$. As
for the INS results, Eq.~(\ref{delta}) can also be generalized to
the case of three sublattices, by substituting $B_{20}$, $\Delta$
and $H_{ex}$ with their weighted averages.\cite{nota2}
$\tilde{\Delta}=c_{(I)}\Delta^{(I)}+c_{(II)}\Delta^{(II)}+c_{(III)}\Delta^{(III)}$
depends on the attribution of each INS transition to one RE
sublattice; the average value $\tilde{\Delta}=21.4~{\rm meV}$ is a
good estimate as far as the abundance of sites II and III in the
crystal is not too different from that of site I.\cite{nota3} (The
latter hypothesis was verified by observing the relative
transition strengths.) The values $H_{ex}^{(I)}=1050\pm 150~{\rm
T}$ and $\tilde{B_{20}}=-12\pm 7~{\rm K}$ were obtained from the
$\tilde{B_{20}}$-vs-$H_{ex}^{(I)}$ curves as derived by INS
[Eq.~(\ref{delta})] and SPD [Eq.~(\ref{k1b20})]. The small value
obtained for $\tilde{B_{20}}$ is in line with the hypothesis made
to interpret the anisotropy data down to 78 K, {\it i.e.} that the
average value results from competing anisotropy contributions of
different sublattices.\cite{alb01,pao02}

In conclusion, the value of the exchange field obtained for
Pr$_{2}$Fe$_{17}$ largely outweighs that of Sm$_{2}$Fe$_{17}$ (380
T)\cite{sol02} as well as those of other common RE-TM compounds
(520 T for Nd$_{2}$Fe$_{14}$B, 270 T for Sm$_{2}$Co$_{17}$, and
450 T for Pr$_{2}$Co$_{17}$).\cite{han93,kuz02,yam88} An
implication of the strong exchange field is that the contributions
of excited $J$ states to the calculated wavefunctions are of the
same order of magnitude than those found in SmFe$_{11}$Ti and in
Sm$_{2}$Co$_{17}$, two RE-TM intermetallics in which the important
role of $J$ mixing was recognized.\cite{moz90,mag00} This is
opposite to what is expected for Pr$^{3+}$ compounds, where
$J$-mixing effects are usually neglected. Moreover, since a large
exchange field strongly enhances anisotropy and reduces the loss
of performance at increasing temperatures, this feature could
potentially make Pr-Fe alloys more interesting than their Nd- and
Sm-based counterparts, assuming that their CF and $T_{C}$ could be
suitably tailored by chemical substitutions.\cite{nota4}


\begin{references}
\bibitem{nota1} This is a good approximation if the exchange field
is almost the same for the three sites, as proved by
Eq.~(\ref{nrt}).
\bibitem{nota2} In principle the numerical coefficient
which multiplies $H_{ex}^{2}$ in the original Eq.~\ref{delta}
should be corrected, but it can be proved that this correction
does not exceed 0.3\%, and is therefore negligible with respect to
other sources of uncertainties which are considered in this work.
\bibitem{nota3} It can be shown that the proposed value of
$\tilde{\Delta}$ is exact in the case that the central INS peak
can be attributed to sublattice I or if the three sites are
equally abundant within the crystal.
\bibitem{nota4} For example, $T_{C}$ can be increased from 282 to
501 K by a 10\% substitution of iron by cobalt
(Ref.~\onlinecite{par02}), and the insertion of small amounts of
interstitial nitrogen or carbon is known to give rise to a strong
easy-axis CF anisotropy in several RE-based compounds.
\end{references}
\end{document}